\documentclass[aps,preprint]{revtex4}
\usepackage{graphicx}
\usepackage{epsfig}
\usepackage{hyperref}
\usepackage{amsmath}

\begin{document}

\title{Revisiting Kerr–Newman–(A)dS Black Holes in \(f(R)\) Gravity: A Kerr–Schild Approach}

\author{\large G\"{o}ksel Daylan Esmer}
\address{Department of Physics, Istanbul University, Vezneciler, 34134 Istanbul, T\"{u}rkiye}
\author{\large Saliha T\"{u}rkmen}
\address{Department of Physics, Istanbul University, Vezneciler, 34134 Istanbul, T\"{u}rkiye}

\begin{abstract}

It is known that, in the constant-curvature regime, the field equations of $f(R)$ gravity reduce to the Einstein form with an effective cosmological constant, establishing a direct correspondence between stationary and axisymmetric black hole solutions in $f(R)$ gravity and their counterparts in general relativity, including the Kerr--Newman--(A)dS family. Within this framework, we revisit the derivation of these metrics using the Kerr--Schild approach for illustrative purposes. The Maxwell equations are solved independently of the Kerr--Schild function, allowing the electromagnetic source to be determined before addressing the gravitational field equations. We then investigate the possibility of non-constant-curvature solutions, taking quadratic $f(R)$ gravity as a representative model. Remarkably, the field equations within the Kerr--Schild ansatz force the Ricci scalar to be constant, reducing the resulting geometries to the familiar Kerr--Newman--(A)dS family. Thus, our analysis extends the recently established no-go result for uncharged black holes to electrically charged configurations within the framework considered here.

\vspace{10mm}

\begin{center}
\textit{Dedicated to beloved \textbf{\textit{Defne}}, with love and wishes for a bright future.}
\end{center}

\end{abstract}

\maketitle
\newpage

\section{Introduction}

Despite its remarkable success, general relativity is not expected to provide a complete description of gravity in all physical regimes, particularly at high curvatures. This limitation has stimulated extensive research into alternative theories of gravity, motivated by theoretical considerations as well as cosmological observations. Among these theories, $f(R)$ gravity has emerged as a widely studied extension in which the Einstein--Hilbert Lagrangian is replaced by a nonlinear function of the Ricci scalar $R$ (see, e.g., Refs.~\cite{Starobinsky:2007hu,Sotiriou:2008rp, {Nojiri:2010wj}}). Despite the higher-order nature of its field equations, the theory offers a relatively tractable framework for exploring gravitational phenomena beyond general relativity.

An important aspect of these investigations concerns black hole solutions, which provide a setting for examining the strong-field behavior of $f(R)$ gravity. Obtaining exact solutions, particularly for rotating configurations, remains a challenging task owing to the complexity of the field equations \cite{{Nojiri:2010wj}, Mignemi:1991wa,delaCruz-Dombriz:2009pzc,Moon:2011hq}. Considerable simplification, however, occurs when the Ricci scalar is constant, $R=R_0$. Under this condition, the field equations take the Einstein form with an effective cosmological constant, establishing a close correspondence with solutions of general relativity. In particular, suitable redefinitions of the metric parameters relate stationary and axisymmetric solutions to the familiar Kerr--(A)dS and Kerr--Newman--(A)dS geometries  \cite{{Cembranos:2011sr},{Larranaga:2011fv}}.
This connection also provides a natural setting in which established methods for constructing rotating black holes can be reconsidered in the context of $f(R)$ gravity.

Among such methods, the Kerr--Schild approach has been particularly influential in the construction of exact rotating black hole solutions, including the Kerr--Newman family in general relativity  \cite{{Kerr:1963ud},{Kerr:1965wfc}}. By expressing the spacetime metric as a deformation of a background geometry along a null vector field, this approach can substantially simplify the Einstein equations, with certain nonlinear contributions admitting an effectively linear description. 
Its significance extends beyond four dimensions, notably through the construction of higher-dimensional Kerr--(A)dS geometries within the generalized Kerr--Schild framework  \cite{{Gibbons:2004js}}. Related formulations have also been employed in certain supergravity theories, including five-dimensional minimal supergravity  \cite{{Aliev:2008bh}}. These developments provide motivation for examining the extent to which the Kerr--Schild construction can accommodate black hole geometries in $f(R)$ gravity, particularly when the assumption of constant scalar curvature is relaxed.

With this motivation, we begin by revisiting the Kerr--Newman--(A)dS solutions in the constant-curvature sector of $f(R)$ gravity, using the Kerr--Schild approach as an illustrative construction. Within this construction, the electromagnetic field is determined directly from the Maxwell equations, independently of the Kerr--Schild function. Having established this connection, we investigate whether the same geometric framework admits rotating and electrically charged solutions beyond the constant-curvature sector. For this purpose, we adopt quadratic $f(R)$ gravity as a representative model and analyze its field equations without assuming $R=R_0$ from the outset. Remarkably, although no constant-curvature condition is imposed a priori, the field equations dynamically enforce a constant Ricci scalar within the Kerr--Schild framework considered here, thereby restricting the resulting solutions to the familiar Kerr--Newman--(A)dS geometries. This finding extends the recently established no-go result for uncharged rotating black holes to electrically charged configurations within the quadratic $f(R)$ gravity model under consideration \cite{Aliev:2026}.

The paper is organized as follows. In Sec. II, we briefly review the field equations of $f(R)$ gravity and revisit the Kerr--Newman--(A)dS solutions in the constant-curvature regime using the Kerr--Schild approach. In Sec. III, we extend the analysis to quadratic $f(R)$ gravity without imposing constant curvature a priori, showing how the field equations dynamically enforce a constant Ricci scalar and select the familiar Kerr--Newman--(A)dS geometries. Finally, Sec. IV is devoted to our conclusions.

\section{Field Equations and the Kerr--Schild Approach}

We consider four-dimensional $f(R)$ gravity minimally coupled to an electromagnetic field, with the dynamics governed by the action \cite{Starobinsky:2007hu,{Nojiri:2010wj}}
\begin{equation}
S=\int d^4x\,\sqrt{-g}\left[\frac{R+f(R)-2\Lambda}{16\pi G}-\frac{1}{16\pi}F_{\mu\nu}F^{\mu\nu}\right],
\label{action}
\end{equation}
where $f(R)$ represents the modification of the Einstein--Hilbert term, $\Lambda$ is the bare cosmological constant, and $F_{\mu\nu}=2\partial_{[\mu}A_{\nu]}$ is the electromagnetic field strength. Independent variations with respect to $g_{\mu\nu}$ and $A_\mu$ give the gravitational and Maxwell equations,
\begin{equation}
(1+f'(R))R_{\mu\nu}-\frac{1}{2}g_{\mu\nu}(R+f(R)-2\Lambda)+(g_{\mu\nu}\nabla^2-\nabla_\mu\nabla_\nu)f'(R)=8\pi G T_{\mu\nu},
\label{eq1}
\end{equation}
\begin{equation}
\nabla_\nu F^{\mu\nu}=0,
\label{eq2}
\end{equation}
where $f'(R)=df/dR$, $\nabla^2=\nabla^\mu\nabla_\mu$, and $T_{\mu\nu}$ denotes the electromagnetic energy--momentum tensor. Since the Maxwell stress tensor is traceless in four dimensions, contraction of Eq.~(\ref{eq1}) leads to
\begin{equation}
3\nabla^2 f'(R)+R(1+f'(R))-2(R+f(R)-2\Lambda)=0.
\label{treq1}
\end{equation}

A particularly simple sector of the theory is obtained by restricting the scalar curvature to a constant value, $R=R_0$. In this case, the derivative term in Eq.~(\ref{treq1}) vanishes, leaving an algebraic condition that determines the admissible constant-curvature branches:
\begin{equation}
R_0=\frac{2[f(R_0)-2\Lambda]}{f'(R_0)-1}.
\label{constantR}
\end{equation}
For a nondegenerate branch satisfying $1+f'(R_0)\neq 0$, the gravitational equations consequently reduce to
\begin{equation}
R_{\mu\nu}=\frac{R_0}{4}g_{\mu\nu}+\frac{8\pi G}{1+f'(R_0)}T_{\mu\nu}.
\label{einform}
\end{equation}
Thus, the constant-curvature sector is governed by equations of the Einstein--Maxwell form, with an effective cosmological constant and gravitational coupling given by
\begin{equation}
\lambda=\frac{R_0}{4},\qquad G_{\mathrm{eff}}=\frac{G}{1+f'(R_0)}.
\label{effcosmcon}
\end{equation}
The modification introduced by $f(R)$ is therefore encoded, in this sector, in the allowed curvature values and the effective coupling to the electromagnetic field.
( For further details, see also  Refs. \cite{{Aliev:2026b},{Esmer:2026tig}}). This correspondence provides the basis for applying the Kerr--Schild construction to the Kerr--Newman--(A)dS family.

To construct rotating charged configurations, we adopt a generalized Kerr--Schild representation of the spacetime metric over a maximally symmetric background \cite{{Gibbons:2004js}}. In this formulation, the full geometry is obtained by adding a term quadratic in a null one-form to the background line element,
\begin{equation}
ds^2=d\bar{s}^2+H\,k\otimes k,
\label{KSform}
\end{equation}
where $H$ is a scalar function and $k=k_\mu dx^\mu$ is null and geodesic with respect to both the background and the full metric. We choose the background to be the (anti-)de Sitter spacetime with curvature determined by the effective cosmological constant $\lambda=R_0/4$.

In the coordinate system $x^\mu=(u,r,\theta,\varphi)$, the null one-form can be expressed as
\begin{equation}
k=\frac{\Delta_\theta}{\Xi}\,du
+\frac{\Sigma}{(1-\lambda r^2/3)(r^2+a^2)}\,dr
-\frac{a\sin^2\theta}{\Xi}\,d\varphi,
\label{koneform}
\end{equation}
where $a$ denotes the rotation parameter and
\begin{equation}
\Sigma=r^2+a^2\cos^2\theta,\qquad
\Xi=1+\frac{\lambda a^2}{3},\qquad
\Delta_\theta=1+\frac{\lambda a^2}{3}\cos^2\theta.
\label{paramets}
\end{equation}
The corresponding Kerr--Schild line element takes the explicit form
\begin{equation}
\begin{aligned}
ds^2={}&-\left(1-\frac{\lambda r^2}{3}\right)\frac{\Delta_\theta}{\Xi}\,du^2
+\frac{\Sigma}{(1-\lambda r^2/3)(r^2+a^2)}\,dr^2
+\frac{\Sigma}{\Delta_\theta}\,d\theta^2\\
&+\frac{(r^2+a^2)\sin^2\theta}{\Xi}\,d\varphi^2\\
&+H\left[
\frac{\Delta_\theta}{\Xi}\,du
+\frac{\Sigma}{(1-\lambda r^2/3)(r^2+a^2)}\,dr
-\frac{a\sin^2\theta}{\Xi}\,d\varphi
\right]^2.
\end{aligned}
\label{metric2}
\end{equation}

A useful property of this representation is that the metric determinant is unaffected by the Kerr--Schild deformation,
\begin{equation}
g=-\frac{\Sigma^2\sin^2\theta}{\Xi^2}.
\label{detg}
\end{equation}
Moreover, the null and geodesic character of $k$ leads to a Ricci scalar that depends linearly on the Kerr--Schild function $H$,
\begin{equation}
R=4\lambda+\partial_r^2H+\frac{2}{\Sigma}
\left(H+2r\,\partial_rH\right).
\label{RicciScalarH}
\end{equation}
This property will be particularly useful in determining the admissible form of $H$ from the constant-curvature condition.

\subsection{Electromagnetic Field}

We now introduce the electromagnetic field through a vector potential aligned with the Kerr--Schild null direction. For a stationary and axisymmetric configuration, we consider the ansatz
\begin{equation}
A_\mu=\frac{\chi(r)}{\Sigma}\,k_\mu,
\label{Aansatz}
\end{equation}
where $\chi(r)$ is an undetermined radial function. This choice is compatible with the structure of the Kerr--Newman electromagnetic potential and allows the Maxwell equations to be treated independently of the explicit form of the Kerr--Schild function $H$.

Indeed, owing to the null structure of the ansatz, the contravariant electromagnetic tensor $F^{\mu\nu}$ is independent of $H$. Since the metric determinant is also unaffected by the Kerr--Schild deformation, the Maxwell equations (\ref{eq2}) can be solved before determining the gravitational field. Substitution of the potential (\ref{Aansatz}) into these equations yields
\begin{equation}
r\chi'(r)-\chi(r)=0,
\label{chieq}
\end{equation}
which admits the solution $\chi(r)=Cr$, with $C$ an integration constant.

The electric charge associated with this configuration is determined by the flux of the electromagnetic field through a closed two-surface,
\begin{equation}
Q_{\mathrm{phys}}=\frac{1}{4\pi}\int_{S^2}\star F=-\frac{C}{\Xi}.
\label{Qstar}
\end{equation}
Defining the charge parameter $Q=\Xi Q_{\mathrm{phys}}$, we obtain
\begin{equation}
\chi(r)=-Qr,
\label{chisol}
\end{equation}
and hence the electromagnetic potential takes the form
\begin{equation}
A=-\frac{Qr}{\Sigma}\,k.
\label{Afinal}
\end{equation}
The corresponding field strength follows directly from $F=dA$. Its components are determined entirely by the background geometry and the charge parameter $Q$, without requiring the explicit solution for $H$. This separation allows us to proceed to the gravitational field equations with the electromagnetic source already specified.

\subsection{Constant-Curvature Kerr--Newman--(A)dS Solutions}

We now consider the constant-curvature sector of the theory, characterized by $R=R_0$. As established in Eqs.~(\ref{constantR})--(\ref{effcosmcon}), the trace equation determines the admissible constant-curvature branches, while
the gravitational field equations reduce to the Einstein--Maxwell form with an effective cosmological constant and gravitational coupling.

The condition $R=R_0=4\lambda$, when applied to the Ricci scalar
(\ref{RicciScalarH}), eliminates the background contribution and yields
\begin{equation}
\partial_r^2H+\frac{2}{\Sigma}\left(H+2r\,\partial_rH\right)=0.
\label{HeqScalar}
\end{equation}
Integrating this equation with respect to $r$, while allowing the integration functions to depend on $\theta$, gives
\begin{equation}
H(r,\theta)=\frac{c_1(\theta)r+c_2(\theta)}{\Sigma}.
\label{Hgeneral}
\end{equation}
The remaining angular dependence is constrained by the tensorial field equations. Since the electromagnetic configuration satisfies $T^r{}_\theta=0$, the corresponding gravitational equation requires $R^r{}_\theta=0$. For the metric under consideration, this component becomes
\begin{equation}
R^r{}_\theta=
\frac{(\Sigma-2r^2)c_1'(\theta)-2r\,c_2'(\theta)}{2\Sigma^2}.
\label{Rrtheta}
\end{equation}
Requiring this expression to vanish identically for arbitrary $r$ implies $c_1'(\theta)=c_2'(\theta)=0$. Thus, both integration functions are constants, and the Kerr--Schild profile is restricted to
\begin{equation}
H(r,\theta)=\frac{c_1r+c_2}{\Sigma}.
\label{Hrestricted}
\end{equation}

The remaining gravitational equations determine the relation between $c_2$ and the electromagnetic charge. To present the relevant components compactly, we introduce
\begin{equation}
\mathcal{X}=(r^2+a^2)\Delta_\theta+
\left(1-\frac{\lambda r^2}{3}\right)a^2\sin^2\theta.
\label{Xdef}
\end{equation}

For the Kerr--Schild metric (\ref{metric2}), the nonvanishing mixed components of the curvature equations are
\begin{equation}
\begin{aligned}
R^u{}_u-\lambda
&=-(R^\varphi{}_\varphi-\lambda)
=\frac{c_2\mathcal{X}}{\Xi\Sigma^3},\\[4pt]
R^r{}_r-\lambda
&=-(R^\theta{}_\theta-\lambda)
=\frac{c_2}{\Sigma^2},\\[4pt]
R^u{}_\varphi
&=-\frac{2ac_2(r^2+a^2)\sin^2\theta}{\Xi\Sigma^3},\\[4pt]
R^\varphi{}_u
&=\frac{2ac_2(1-\lambda r^2/3)\Delta_\theta}{\Xi\Sigma^3}.
\end{aligned}
\label{RicciMixed}
\end{equation}
The corresponding electromagnetic energy--momentum tensor components are
\begin{equation}
\begin{aligned}
T^u{}_u&=-T^\varphi{}_\varphi
=-\frac{Q^2\mathcal{X}}{8\pi\Xi\Sigma^3},\\[4pt]
T^r{}_r&=-T^\theta{}_\theta
=-\frac{Q^2}{8\pi\Sigma^2},\\[4pt]
T^u{}_\varphi
&=\frac{aQ^2(r^2+a^2)\sin^2\theta}{4\pi\Xi\Sigma^3},\\[4pt]
T^\varphi{}_u
&=-\frac{aQ^2(1-\lambda r^2/3)\Delta_\theta}{4\pi\Xi\Sigma^3}.
\end{aligned}
\label{MaxwellMixed}
\end{equation}
The remaining components of the field equations vanish identically. Substitution into (\ref{einform}) shows that all independent equations consistently impose the same algebraic condition,
\begin{equation}
c_2=-q^2,\qquad
q^2=\frac{GQ^2}{1+f'(R_0)}.
\label{qfunc}
\end{equation}
We assume $1+f'(R_0)>0$, corresponding to a positive effective gravitational coupling.

The integration constant $c_1$ determines the leading mass contribution to the geometry. With the conventional parametrization $c_1=2M$, where $M$ denotes the mass parameter, the Kerr--Schild function takes the form
\begin{equation}
H(r,\theta)=\frac{2Mr-q^2}{\Sigma}.
\label{Hsol}
\end{equation}
This is precisely the radial structure associated with the Kerr--Newman--(A)dS family, with the charge contribution rescaled by the effective gravitational coupling.

To recover the familiar Boyer--Lindquist representation, we introduce the coordinate transformations
\begin{equation}
\begin{aligned}
du &= dt+\frac{2Mr-q^2}{(1-\lambda r^2/3)\Delta_r}\,dr,\\[4pt]
d\varphi &= d\phi-\frac{\lambda a}{3}\,dt
+\frac{a(2Mr-q^2)}{(r^2+a^2)\Delta_r}\,dr,
\end{aligned}
\label{coordtrans}
\end{equation}
where
\begin{equation}
\Delta_r=(r^2+a^2)\left(1-\frac{\lambda r^2}{3}\right)-2Mr+q^2.
\label{Deltar}
\end{equation}
In these coordinates, the line element becomes
\begin{equation}
\begin{aligned}
ds^2={}&-\frac{\Delta_r}{\Sigma}
\left(dt-\frac{a\sin^2\theta}{\Xi}d\phi\right)^2
+\frac{\Sigma}{\Delta_r}dr^2
+\frac{\Sigma}{\Delta_\theta}d\theta^2\\
&+\frac{\Delta_\theta\sin^2\theta}{\Sigma}
\left(a\,dt-\frac{r^2+a^2}{\Xi}d\phi\right)^2.
\end{aligned}
\label{metric}
\end{equation}
We thus recover the Kerr--Newman--(A)dS geometry in the constant-curvature sector of $f(R)$ gravity. Notably, the equation governing the radial dependence of $H$ is independent of the explicit functional form of $f(R)$, whose influence is encoded in the allowed effective cosmological constant and the rescaled charge parameter. This construction provides an illustrative realization of the correspondence with Einstein--Maxwell gravity and establishes the geometric framework for investigating non-constant-curvature configurations in the following section.

\section{Arbitrary Curvature: Quadratic Model}

We now investigate whether the Kerr--Schild construction developed in the preceding section can support rotating and electrically charged configurations with a non-constant Ricci scalar. To address this question, we consider the quadratic modification
\begin{equation}
f(R)=\alpha R^2,
\label{quadf}
\end{equation}
where $\alpha$ is the coupling constant of the quadratic curvature term. Accordingly, the gravitational Lagrangian in the action (\ref{action}) becomes $R+\alpha R^2-2\Lambda$. The field equations (\ref{eq1}) then read
\begin{equation}
\begin{aligned}
E_{\mu\nu}={}&(1+2\alpha R)R_{\mu\nu}
-\frac{1}{2}(R+\alpha R^2-2\Lambda)g_{\mu\nu}\\
&+2\alpha(g_{\mu\nu}\nabla^2-\nabla_\mu\nabla_\nu)R
=8\pi G T_{\mu\nu}.
\end{aligned}
\label{quadEq}
\end{equation}
Taking the trace and using the tracelessness of the electromagnetic energy--momentum tensor gives
\begin{equation}
6\alpha\nabla^2R-R+4\Lambda=0.
\label{quadtrace}
\end{equation}

To investigate possible deviations from the Kerr--Newman--(A)dS family, we allow the radial part of the Kerr--Schild deformation to remain unspecified, while preserving the angular structure inherited from the rotating background \cite{Aliev:2026}.

\begin{equation}
H(r,\theta)=\frac{m(r)}{\Sigma},
\label{Hm}
\end{equation}

This choice is motivated by the constant-curvature solution obtained in the preceding section, where the dependence on the angular coordinate enters through $\Sigma$. By promoting the radial numerator to an unknown function $m(r)$, we can examine whether the higher-curvature contributions generate additional radial behavior beyond the Einstein--Maxwell branch. The function $m(r)$ will be determined solely by the modified gravitational field equations, without imposing constant scalar curvature in advance.

The electromagnetic potential remains that determined by Eqs.~(\ref{Aansatz}) and (\ref{chisol}), since the Maxwell equations (\ref{eq2}) do not depend on the Kerr--Schild function $H$. Substituting the ansatz (\ref{Hm}) into the Ricci scalar (\ref{RicciScalarH}), we obtain
\begin{equation}
R-R_0=\frac{m''(r)}{\Sigma},
\label{Rm}
\end{equation}
where $R_0=4\lambda$ is the scalar curvature of the background geometry.

We next examine the off-diagonal components of the quadratic field equations (\ref{quadEq}). The $r\theta$ component is particularly useful because the electromagnetic stress tensor satisfies $T_{r\theta}=0$, so that this component provides a purely geometric constraint. Substituting the curvature relation (\ref{Rm}) into the field equations, we obtain
\begin{equation}
E_{r\theta}=-\frac{3\alpha a^2\sin 2\theta}{\Sigma^3}\left[\Sigma m'''(r)-4r m''(r)\right]=0.
\label{Ertheta}
\end{equation}

For a rotating configuration with $a\neq 0$ and a nonvanishing quadratic coupling $\alpha\neq 0$, this equation requires
\begin{equation}
\Sigma m'''(r)-4r m''(r)=0.
\label{rthetacond}
\end{equation}
Since $m(r)$ depends only on the radial coordinate, whereas $\Sigma=r^2+a^2\cos^2\theta$ contains explicit angular dependence, Eq.~(\ref{rthetacond}) must hold for arbitrary values of $\theta$. For $a\neq 0$, this requires $m'''(r)=0$. Substituting this result back into Eq.~(\ref{rthetacond}) then gives $m''(r)=0$.

Substituting these conditions into the trace equation (\ref{quadtrace}) confirms the vanishing of the fourth radial derivative, $m^{(4)}(r)=0$, consistently with the preceding constraints. We therefore obtain
\begin{equation}
m''(r)=m'''(r)=m^{(4)}(r)=0.
\label{conds}
\end{equation}
With the consistency conditions (\ref{conds}) imposed, we now examine the remaining components of the gravitational field equations in the presence of the electromagnetic source. It is convenient to introduce the radial combination
\begin{equation}
K(r)=r\,m'(r)-m(r).
\label{Kdef}
\end{equation}
The relevant off-diagonal components of the quadratic gravitational equations can then be expressed as
\begin{equation}
\begin{aligned}
E_{ur}&=-\frac{(1+8\alpha\Lambda)\Delta_\theta\,m(r)}{\Xi(1-\lambda r^2/3)(r^2+a^2)\Sigma^2}\,K(r),
\qquad
E_{r\varphi}=-\frac{a\sin^2\theta}{\Delta_\theta}\,E_{ur},\\[6pt]
E_{u\varphi}&=\frac{(1+8\alpha\Lambda)a\sin^2\theta\,\Delta_\theta\left[m(r)-2(r^2+a^2)(1-\lambda r^2/3)\right]}{\Xi^2\Sigma^3}\,K(r).
\end{aligned}
\label{Eoff}
\end{equation}
For the electromagnetic field specified in Sec.~II, the corresponding components of the energy--momentum tensor are
\begin{equation}
\begin{aligned}
T_{ur}&=-\frac{Q^2\Delta_\theta\,m(r)}{8\pi\Xi(1-\lambda r^2/3)(r^2+a^2)\Sigma^2},
\qquad
T_{r\varphi}=-\frac{a\sin^2\theta}{\Delta_\theta}\,T_{ur},\\[6pt]
T_{u\varphi}&=\frac{Q^2a\sin^2\theta\,\Delta_\theta\left[m(r)-2(r^2+a^2)(1-\lambda r^2/3)\right]}{8\pi\Xi^2\Sigma^3}.
\end{aligned}
\label{Toff}
\end{equation}
Substituting Eqs.~(\ref{Eoff}) and (\ref{Toff}) into the field equations (\ref{quadEq}), we find that all independent off-diagonal components lead to a single differential equation for $m(r)$. The diagonal components yield no additional independent constraints. Thus, the remaining field equations reduce to
\begin{equation}
r\,m'(r)-m(r)=q^2,\qquad q^2=\frac{GQ^2}{1+8\alpha\Lambda}.
\label{Keq}
\end{equation}
Here we assume $1+8\alpha\Lambda>0$, corresponding to a positive effective gravitational coupling, consistently with the constant-curvature sector discussed in Sec.~II.

The first-order differential equation (\ref{Keq}) admits the general solution
\begin{equation}
m(r)=Cr-q^2,
\label{msol}
\end{equation}
where $C$ is an integration constant. Identifying $C=2M$, with $M$ denoting the mass parameter, the Kerr--Schild function (\ref{Hm}) becomes
\begin{equation}
H(r,\theta)=\frac{2Mr-q^2}{\Sigma}.
\label{Hquad}
\end{equation}
The resulting Kerr--Schild function (\ref{Hquad}) coincides with that of the Kerr--Newman--(A)dS family obtained in Sec.~II, with the electromagnetic charge rescaled by the effective gravitational coupling. Remarkably, although no constant-curvature condition was imposed a priori, the quadratic field equations dynamically force the Ricci scalar to become constant and uniquely select the Kerr--Newman--(A)dS geometry within the class of rotating and electrically charged Kerr--Schild configurations considered here. Consequently, no additional non-Einstein solutions arise within this class, extending the corresponding no-go result \cite{Aliev:2026} for uncharged rotating black holes to electrically charged configurations.

\section{Discussion}

In this work, we have revisited the Kerr--Newman--(A)dS black hole solutions in $f(R)$ gravity using the generalized Kerr--Schild approach and investigated the possibility of rotating and electrically charged configurations beyond the constant-curvature sector. An important aspect of our construction is that the electromagnetic field is determined directly from the Maxwell equations by adopting a vector potential aligned with the Kerr--Schild null direction. Owing to the geometric properties of this ansatz, the Maxwell equations can be solved independently of the unknown Kerr--Schild function, allowing the electromagnetic source to be specified before solving the gravitational field equations.

In the constant-curvature regime, the modified gravitational equations reduce to the Einstein--Maxwell form with an effective cosmological constant and gravitational coupling. Within the Kerr--Schild framework, we explicitly recover the Kerr--Newman--(A)dS geometry, with the effects of $f(R)$ gravity encoded in the allowed curvature branches and the rescaled electromagnetic charge. This construction provides a direct illustration of the correspondence between constant-curvature $f(R)$ solutions and their general relativistic counterparts.

The central result emerges in quadratic $f(R)$ gravity when the constant-curvature condition is not imposed a priori. Allowing an arbitrary radial deformation of the Kerr--Schild profile, we find that the off-diagonal field equations dynamically force the Ricci scalar to become constant. The remaining equations then uniquely determine the radial profile, recovering the Kerr--Newman--(A)dS family. Remarkably, the inclusion of electric charge does not relax these constraints, but merely modifies the effective charge parameter. Our analysis thus extends the no-go result for uncharged rotating configurations to the electrically charged sector.

Our results show that quadratic curvature corrections do not necessarily give rise to new rotating black hole geometries, even in the presence of an electromagnetic field. While the present analysis is restricted to the Kerr--Schild class considered here, it reveals how the combined requirements of rotation and the modified gravitational field equations severely constrain possible departures from the Einstein--Maxwell branch. In particular, the inclusion of electric charge does not circumvent these restrictions, reinforcing the no-go result within the class of geometries under consideration.

\section*{Acknowledgments}
The authors wish to thank A. N. Aliev for suggesting this problem and for very helpful discussions.


\begin{thebibliography}{99}

\bibitem{Starobinsky:2007hu} A. A. Starobinsky, JETP Lett. {\bf 86}, 157 (2007).

\bibitem{Sotiriou:2008rp} T. P. Sotiriou and V. Faraoni, Rev. Mod. Phys. {\bf 82}, 451 (2010).

\bibitem{Nojiri:2010wj} S. Nojiri and S. D. Odintsov, Phys. Rept. {\bf 505}, 59 (2011).

\bibitem{Mignemi:1991wa} S. Mignemi and D. L. Wiltshire, Phys. Rev. D {\bf 46}, 1475 (1992).

\bibitem{delaCruz-Dombriz:2009pzc} A. de la Cruz-Dombriz, A. Dobado and A. L. Maroto, Phys. Rev. D {\bf 80}, 124011 (2009) [Erratum: Phys. Rev. D {\bf 83}, 029903 (2011)].

\bibitem{Moon:2011hq} T. Moon, Y. S. Myung and E. J. Son, Gen. Rel. Grav. {\bf 43}, 3079 (2011).

\bibitem{Cembranos:2011sr} J. A. R. Cembranos, A. de la Cruz-Dombriz and P. Jimeno Romero, Int. J. Geom. Methods Mod. Phys. {\bf 11}, 1450001 (2014).

\bibitem{Larranaga:2011fv} A. Larranaga, Pramana {\bf 78}, 697 (2012).

\bibitem{Kerr:1963ud} R. P. Kerr, Phys. Rev. Lett. {\bf 11}, 237 (1963).

\bibitem{Kerr:1965wfc} R. P. Kerr and A. Schild, Proc. Symp. Appl. Math. {\bf 17}, 199 (1965).

\bibitem{Gibbons:2004js} G. W. Gibbons, H. L\"u, D. N. Page and C. N. Pope, Phys. Rev. Lett. {\bf 93}, 171102 (2004).

\bibitem{Aliev:2008bh} A. N. Aliev and D. K. Çiftçi, Phys. Rev. D {\bf 79}, 044004 (2009).

\bibitem{Aliev:2026} A. N. Aliev, arXiv:2606.02929 [gr-qc].

\bibitem{Aliev:2026b} A. N. Aliev and G. D. Esmer, Phys. Rev. D {\bf 113}, 044074 (2026).

\bibitem{Esmer:2026tig} G. D. Esmer and S. T\"urkmen, Int. J. Geom. Methods Mod. Phys. {\bf 23}, 2650082 (2026).

\end{thebibliography}
\end{document}